\documentclass[twocolumn,showpacs,superscriptaddress,10pt]{revtex4-1} 
\usepackage{amsmath,amssymb}
\usepackage{graphicx}
\usepackage{color}
\begin{document}

\renewcommand{\deg}{^{\circ}}
\newcommand{\micron}{\mu\mbox{m}}

\newcommand{\WASEDA}{\affiliation{Department of Applied Physics,
    School of Advanced Science and Engineering, Waseda University,
    3-4-1 Okubo, Shinjuku-ku, Tokyo 169-8555, Japan}}
\newcommand{\KANAZAWA}{\affiliation{Faculty of Mechanical Engineering,
    Institute of Science and Engineering, Kanazawa University,
    Kakuma-machi, Kanazawa, Ishikawa 920-1192, Japan}}
\title{Self-adjustment of a nonlinear lasing mode to a pumped area in a two-dimensional microcavity}

\author{Yuta Kawashima}\WASEDA
\author{Susumu Shinohara}\WASEDA
\author{Satoshi Sunada}\KANAZAWA
\author{Takahisa Harayama}\WASEDA

\begin{abstract}
We numerically performed wave dynamical simulations based on the Maxwell-Bloch (MB) model for a quadrupole-deformed microcavity laser with spatially selective pumping. We demonstrate the appearance of an asymmetric lasing mode whose spatial pattern violates both the $x$- and $y$-axes mirror symmetries of the cavity. Dynamical simulations revealed that a lasing mode consisting of a clockwise or counterclockwise rotating-wave component is a stable stationary solution of the MB model. From the results of a passive-cavity mode analysis, we interpret these asymmetric rotating-wave lasing modes by the locking of four nearly degenerate passive-cavity modes. For comparison, we carried out simulations for a uniform pumping case and found a different locking rule for the nearly degenerate modes. Our results demonstrate a nonlinear dynamical mechanism for the formation of a lasing mode that adjusts its pattern to a pumped area.

\vspace{3mm}
\vskip 5mm
\noindent
\href{https://doi.org/10.1364/PRJ.5.000B47}{doi:10.1364/PRJ.5.000B47}
\vskip 3mm
\noindent
\copyright~2017~Optical Society of America. Users may use, reuse, and
build upon the article, or use the article for text or data mining, so
long as such uses are for non-commercial purposes and appropriate
attribution is maintained. All other rights are reserved.

\end{abstract}
\maketitle 

\section{INTRODUCTION}
Since a close analogy between optical microcavities and open dynamical
billiards was pointed out \cite{Noeckel97}, intensive investigations
have been carried out for ray-chaotic optical cavities, whose typical
shape is a deformation of a circle \cite{Schwefel04a, Harayama11,
  Cao15, Jiang16}.
Combination of total internal reflection and ray chaos is expected to
achieve a low-threshold and directional-emission microcavity laser
\cite{Wiersig08, Wiersig11}.

Recently, Aung et al. experimentally demonstrated that threshold
current reduction and directional emission can be simultaneously
achieved by forming an appropriate selective pumping area in the
quadrupole-deformed microcavity laser \cite{Aung15}.
In this cavity, there are resonant modes which are localized along
twin periodic orbits with the shape of the double triangle consisting
of upward-pointing and downward-pointing triangles [see
  Fig. \ref{fig:quadrupole}(a)].
In the experiments reported in Ref. \cite{Aung15}, the selective
pumping area was formed along one of the two triangle orbits, and
directional emission was observed.
The assumption of the existence of a lasing mode that localizes along
the sole triangle orbit plays a key role in explaining the
experimental data in Ref. \cite{Aung15}.
With this assumption, the directional emission can be explained by the
mechanism of unstable-manifold-guided emission \cite{Schwefel04b,
  Lee05, Shinohara06} via dynamical tunneling \cite{Podolskiy05,
  Shinohara10, Yang10}, and the threshold current reduction can be
attributed to the fact that the triangle orbit is confined by total
internal reflection.
However, because this lasing mode violates the mirror symmetry of the
quadrupole cavity, it cannot be simply regarded as a passive-cavity
mode.

A theoretical explanation is needed on how the selective pumping leads
to the formation of the asymmetric lasing mode, which does not have a
direct counterpart in the passive-cavity modes.
It is the purpose of this paper to numerically demonstrate the
existence of the asymmetric lasing mode and theoretically reveal its
appearance mechanism.
We perform wave dynamical simulations based on the Maxwell-Bloch (MB)
model \cite{Harayama05, Harayama11}, which fully takes into account
the nonlinear interaction between the light field and a gain medium
described by a two-level atom system (i.e., the optical Bloch
equations \cite{Loudon00}).
We numerically reproduce the appearance of the asymmetric lasing mode
when the selective pumping condition is adopted, and explain it by the
locking of four nearly degenerate modes associated with the
double-triangle orbits.
Dynamical simulations reveal that clockwise (CW) and counterclockwise
(CCW) rotating-wave states are stable stationary solutions of the MB
model.
For comparison, we perform simulations for a uniform pumping case, and
found a different locking rule for the nearly degenerate modes.

\section{MODEL}
\subsection{MB Model}
The MB model is a set of wave dynamical equations describing the
nonlinear interaction between the light field and a gain medium
\cite{Harayama05, Tureci06, Harayama11}.
For transverse-magnetic (TM) polarization, it is given by
\begin{align}
  &\frac{\partial^2}{\partial t^2} \left(E_z+\frac{4\pi}{\epsilon}P_z\right)=
  \frac{c^2}{n^2}\nabla^2 E_z-2\beta\frac{\partial}{\partial t} E_z,\label{eq:MB1}\\
  &P_z=N(\rho+\rho^*)\kappa\hbar,\label{eq:MB2}\\
  &\frac{\partial}{\partial t}\rho=-i\omega_0\rho-i\kappa W E_z-\gamma_{\perp}\rho,\label{eq:MB3}\\
  &\frac{\partial}{\partial t}W=-2i\kappa
  E_z(\rho-\rho^*)-\gamma_{\parallel}(W-W_{\infty}),\label{eq:MB4}
\end{align}
where $E_z(x,y)$, $P_z(x,y)$, and $W(x,y)$ are the electric field,
polarization, and population inversion, respectively.
$n$ $=$ $n(x,y)$ is the refractive index, and $\epsilon$ $=$ $n^2$ is
the permittivity.
The constant $\beta$ is introduced phenomenologically to describe the
background uniform absorption.
$\rho(x,y)$ is the microscopic polarization, $N$ the atomic number
density, $\kappa$ the coupling strength for the light-matter
interaction, $\omega_0$ the transition frequency of the two-level gain
medium, and $W_{\infty}(x,y)$ the pumping strength parameter.
$\gamma_{\perp}$ and $\gamma_{\parallel}$ are the transversal and
longitudinal relaxation rates, respectively.
In the study reported in this paper, we numerically solve
Eq. (\ref{eq:MB1}) by the finite-difference time-domain method and
Eqs. (\ref{eq:MB3}) and (\ref{eq:MB4}) by the Euler method.

\subsection{The Quadrupole-Deformed Cavity}
\begin{figure}[b!]
\centerline{\includegraphics[width=1.0\columnwidth]{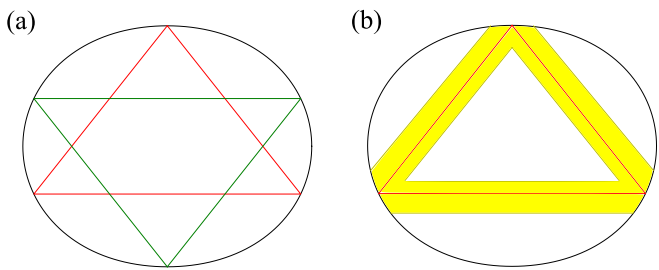}}
\caption{(a) Double-triangle orbits in the quadrupole-deformed
  cavity. (b) Spatial selective pumping (yellow region) along the
  upward-pointing triangle orbit (red lines).}
\label{fig:quadrupole}
\end{figure}
In the MB model, the effect of the cavity shape appears through the
refractive index function $n(x,y)$.
We fix the cavity shape to be the quadrupole-deformed cavity defined
in the polar coordinates $(r,\theta)$ by
\begin{equation}
r(\theta)=r_0 \left[1+\varepsilon\cos(2\theta)\right],
\end{equation}
where $r_0$ is the size parameter, and the deformation parameter
$\varepsilon$ is fixed at $0.09$ throughout the paper.
The refractive indices inside and outside the cavity are fixed at
$n_{in}$ $=$ $3.3$ (GaAs) and $n_{out}$ $=$ 1 (air), respectively.
In this paper, we focus our attention on the periodic orbits with the
shape of the double triangle, consisting of upward-pointing and
downward-pointing stable triangle orbits as shown in
Fig. \ref{fig:quadrupole}(a).

In the experiments by Aung et al. \cite{Aung15}, the effective
refractive index of the cavity is 3.67, and the emitted light is
transverse-electric polarized, which are different from our setting.
However, we believe that our claims presented in this paper are
qualitatively applicable to the microcavity laser studied by Aung et
al.\\

\subsection{Resonant Modes for the Passive Cavity}
\label{sect:resonant_modes}

\begin{figure}[b!]
\centerline{\includegraphics[width=1.0\columnwidth]{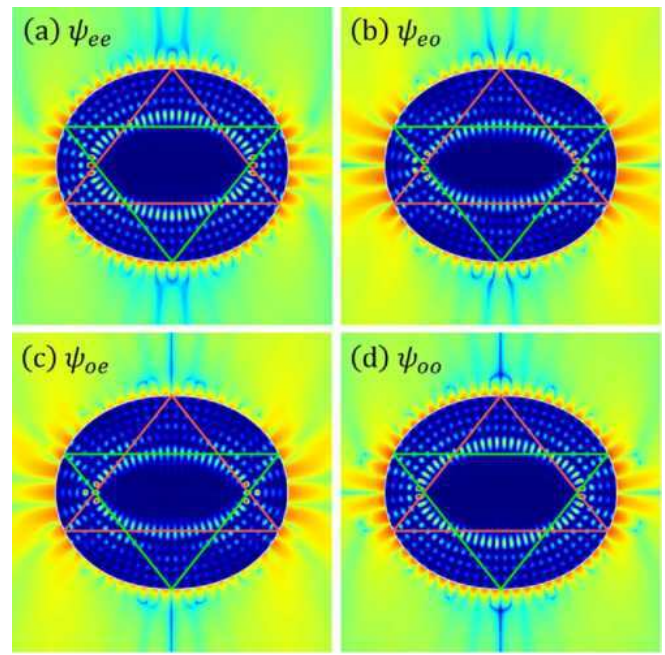}}
\caption{Intensity distributions of the resonant modes for a passive
  quadrupole-deformed cavity with refractive index 3.3. The modes are
  four nearly degenerate modes associated with the double-triangle
  orbits.  The double-triangle orbits (red and green lines) are
  superposed, and the intensities outside the cavity are plotted in
  log scale. (a)~Even-even mode with scaled frequency
  $\mbox{Re}\,\omega/\omega_0$ $=$ $1.0008278$. (b) Even-odd mode with
  $\mbox{Re}\,\omega/\omega_0$ $=$ $0.999085$. (c) Odd-even mode with
  $\mbox{Re}\,\omega/\omega_0$ $=$ $0.999075$. (d) Odd-odd mode with
  $\mbox{Re}\,\omega/\omega_0$ $=$ $1.0008277$.}
\label{fig:wf_resonances}
\end{figure}
\begin{figure}[t!]
\centerline{\includegraphics[width=1.0\columnwidth]{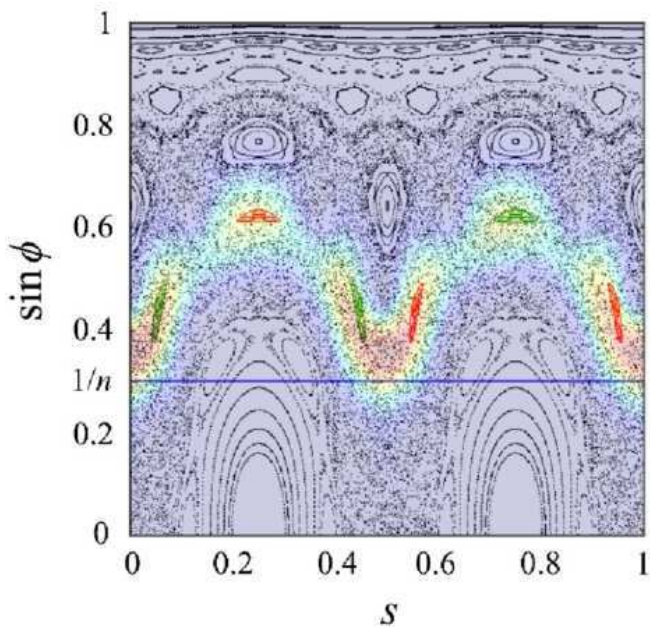}}
\caption{Phase space of the ray dynamics for the quadrupole-deformed
  cavity. The islands of stability corresponding to the
  upward-pointing and downward-pointing triangle orbits are
  indicated by red and green points, respectively. The critical line
  for total internal reflection is indicated by a line at $\sin\phi$
  $=$ $1/3.3$. Husimi distribution for the $eo$ mode shown in
  Fig. \ref{fig:wf_resonances} (b) is superposed.}
\label{fig:husimi}
\end{figure}

The resonant modes for the passive cavity are the solutions of
Eq. (\ref{eq:MB1}) with $P_z$ $\equiv$ $0$ and $\beta$ $=$ $0$.
That is, they are the eigensolutions of the following Helmholtz
equation:
\begin{equation}
\left(\nabla^2 + \frac{n^2 \omega^2}{c^2}\right) \psi(x,y)=0,
\label{eq:Helmholtz}
\end{equation}
where $E_z$ $=$ $\mbox{Re}\,\left[\psi(x,y)\exp(-i\omega t)\right]$.
Because we assume TM polarization, we impose that both the wave
function and its normal derivative are continuous at the cavity
boundary.
The eigenfrequency $\omega$ takes a complex value with a negative
imaginary part, because Eq. (\ref{eq:Helmholtz}) is solved with the
outgoing wave condition at infinity, i.e., $\psi\propto e^{i\omega
  r/c}/\sqrt{r}$ as $r\to\infty$.
Because the quadrupole cavity has mirror symmetries with respect to
both the $x$ and $y$ axes, the resonant modes are divided into four
symmetry classes, i.e.,
\begin{align}
\psi_{ab}(-x,y) &= a\,\psi_{ab}(x,y),\\
\psi_{ab}(x,-y) &= b\,\psi_{ab}(x,y),
\end{align}
where $a$,$b$ $\in$ $\{+,-\}$ are parity indices.

For the double-triangle orbits, the associated resonant modes can be
shown to have fourfold near degeneracy by a symmetry argument
\cite{Tureci02}.
Figure \ref{fig:wf_resonances} shows an example of the four nearly
degenerate modes with $\mbox{Re}\,\omega/\omega_0$ $\approx$ $1$,
where the even-even ($ee$) and odd-odd ($oo$) modes constitute a
closer pair, whereas the even-odd ($eo$) and odd-even ($oe$) modes form
another closer pair.
Here we used the scaled eigenfrequency $\mbox{Re}\,\omega/\omega_0$,
where $\omega_0$ $=$ $2.2$ $\times$ $10^{15}$ is the transition
frequency of the two-level gain medium tuned to fit the frequencies of
the nearly degenerate modes.
We numerically obtained the eigenfrequencies and eigenfunctions by the
boundary element method \cite{Wiersig03}.

Because of the relatively small wavenumbers, the intensity
localization along the triangle orbits is less obvious in
Fig. \ref{fig:wf_resonances}.
However, by investigating the Husimi distributions \cite{Hentschel03}
of the wave functions, we could identify the intensity localization
along the double-triangle orbits.
Figure \ref{fig:husimi} shows the upper-half phase space for the ray
dynamics \cite{Schwefel04a, Harayama11, Cao15}, where the phase space
is spanned by $(s,\sin\phi)$ with $s$ and $\phi$ being the arc length
along the cavity boundary and the incident angle for a ray orbit,
respectively.
In Fig. \ref{fig:husimi}, the islands of stability corresponding to
the upward-pointing and downward-pointing triangle orbits are shown
by red and green points, respectively.
In Fig. \ref{fig:husimi}, we also superpose the Husimi distribution
for the resonant mode $eo$, whose wave function pattern is shown in
Fig. \ref{fig:wf_resonances} (b).
In the Husimi distribution, we can see high intensities near the
islands of stability.
We also checked that modes similar to those in
Fig. \ref{fig:wf_resonances} appear regularly in the $\omega$ plane
with a constant modal spacing corresponding the optical path length of
the triangle orbit.

As can be seen in Fig. \ref{fig:husimi}, the double-triangle orbits
are located above the critical line for total internal reflection.
The emission of a well-confined mode in a ray-chaotic cavity can be
explained by the chaos-assisted emission (CAE) mechanism
\cite{Podolskiy05, Shinohara10, Yang10}.
As Aung et al. have shown \cite{Aung15}, the CAE mechanism for the
quadrupole cavity results in strong emissions at the far-field angles
$\theta$ $\approx$ $50\deg$, $130\deg$, $230\deg$, and $310\deg$.

\begin{figure}[t!]
\centerline{\includegraphics[width=1.0\columnwidth]{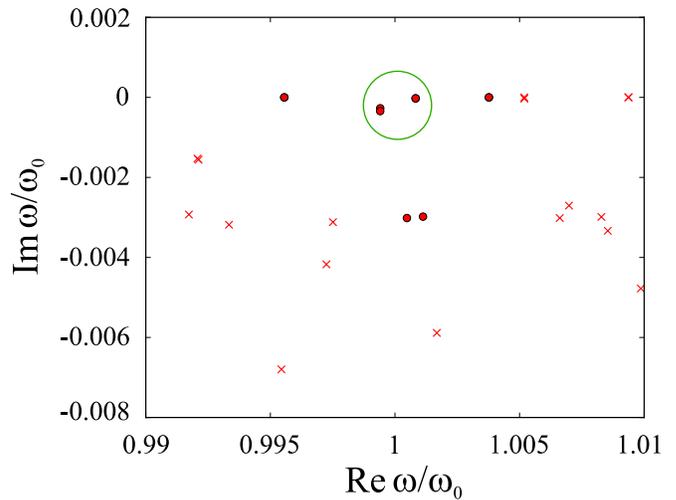}}
\caption{Distribution of the complex eigenfrequencies $\omega$ scaled
  by $\omega_0$, where $\omega_0$ is the gain center parameter. The
  four nearly degenerate modes associated with the double-triangle
  orbits are encircled by a green circle (the $ee$ and $oo$ modes are
  almost on top of each other, and so are the $eo$ and $oe$ modes).
The modes indicated by filled circles ($\bullet$) have positive linear
gain [i.e., satisfying Eq. (\ref{eq:positive_gain_condition})] for the
selective pumping with $W_{\infty}$ $=$ $1.0$ $\times$ $10^{-3}$,
whereas those indicated by crosses ($\times$) do not satisfy
Eq. (\ref{eq:positive_gain_condition}).}
\label{fig:resonances_dist}
\end{figure}

Figure \ref{fig:resonances_dist} shows the eigenfrequency distribution
of the resonant modes in the complex $\omega/\omega_0$ plane, where Im
$\omega/\omega_0$ represents the decay rate of the mode.
The modes encircled by the green circle are the four nearly degenerate
modes associated with the double-triangle orbits (the $ee$ and $oo$
modes are almost on top each other, and so are the $eo$ and $oe$
modes).
The longitudinal modal spacing for the double-triangle modes is
estimated to be $(\Delta \mbox{Re}\,\omega)/\omega_0 \approx 0.0244$
from the optical path length of the triangle orbit.
For the numerical simulations reported in this paper, we only consider
the cases where a single set of the four nearly degenerate modes has
positive gain.

In the eigenfrequency distribution, we can see that some modes are
aligned at Im\,$(\omega/\omega_0)$ $\approx$ $-0.003$.
This decay rate value was found to agree with that estimated by the
Fresnel-coefficient-weighted ray simulation \cite{Harayama11} for
chaotic orbits located near the critical line for total internal
reflection in the phase space.
Thus, these modes are associated with the chaotic orbits.\\

\section{NUMERICAL RESULTS OF THE MB MODEL SIMULATION}
\subsection{Positive Linear Gain Condition for a Resonant Mode}
The condition for a resonant mode to have positive linear gain is
derived in the limit of a single-mode approximation ignoring modal
couplings \cite{Harayama05}:
\begin{equation}
\frac{2\pi N\kappa^2\hbar\,\langle W_{\infty} \rangle}{n^2}
\frac{\gamma_{\perp}\,\mbox{Re}\,\omega_s}{\left(\mbox{Re}\,\omega_s-\omega_0\right)^2+\gamma_{\perp}^2}
>-\mbox{Im}\,\omega_s+\beta,
\label{eq:positive_gain_condition}
\end{equation}
with
\begin{equation}
\langle W_{\infty} \rangle=\frac{W_{\infty} \int_{\cal D} dxdy\,
  |\psi(x,y)|^2 \Theta(x,y)}{\int_{\cal D} dxdy\, |\psi(x,y)|^2},
\label{eq:W_inf}
\end{equation}
where $\omega_s$ $(s\in\mathbb{N})$ is the eigenfrequency of a
passive-cavity mode, ${\cal D}$ denotes the area inside the cavity,
and $\Theta(x,y)$ represents a characteristic function that takes $1$
inside the pumped area (i.e., the area along the upward-pointing
triangle orbit) and takes $0$ otherwise.
We note that this condition depends on the wave function $\psi(x,y)$
through Eq. (\ref{eq:W_inf}), and that the gain center is controlled
by the parameter $\omega_0$.
This condition is useful for estimating which passive-cavity modes
have the potential to contribute to a lasing state for given gain
center $\omega_0$ and pumping strength $W_{\infty}$.

For the simulations reported in this paper, we fixed the parameter
values as follows: $r_0$ $=$ $2.027$ $\micron$, $\beta$ $=$ $8.8$
$\times$ $10^{12}$ s$^{-1}$, $\omega_0$ $=$ $2.2$ $\times$ $10^{15}$
s$^{-1}$, $\gamma_{\perp}$ $=$ $13.2$ $\times$ $10^{12}$ s$^{-1}$,
$\gamma_{\parallel}$ $=$ $6.6$ $\times$ $10^{12}$ s$^{-1}$, and
$N\kappa^2\hbar$ $=$ $0.55$ J s$^{-1}$ cm$^{-3}$.
As mentioned in Section \ref{sect:resonant_modes}, the value of
$\omega_0$ is chosen so that the double-triangle modes become the
nearest to the gain center.
The pumping strength parameter $W_{\infty}$ is dimensionless, and it
is varied in the simulations.

\subsection{Selective Pumping Simulations}
\begin{figure}[t!]
\centerline{\includegraphics[width=1.0\columnwidth]{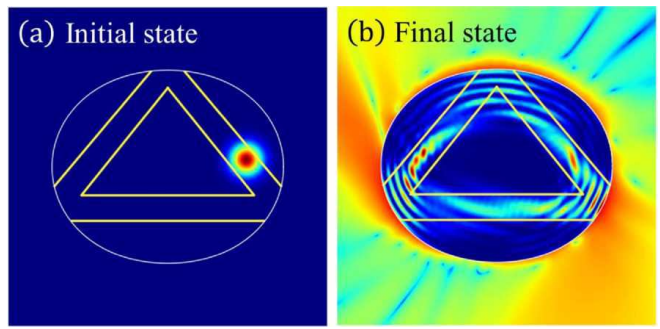}}
\caption{Electric field intensity distributions. (a) An initial
  condition for the MB model simulation. (b) Time-averaged pattern of
  the stationary lasing state of the MB model for the selective
  pumping case with $W_{\infty}$ $=$ $1.0$ $\times$ $10^{-3}$. The
  intensity outside the cavity is plotted in log scale. The boundary
  of the pumped area is indicated by yellow lines.}
\label{fig:MB_states}
\end{figure}

Selective pumping in two-dimensional microcavity lasers has been
experimentally demonstrated in Refs. \cite{Aung15,Shinohara10,
  Fukushima02, Chern03, Fukushima04, Choi06, Liew15}, and
theoretically studied in Refs. \cite{Deych05, Kwon06, Ge10, Ge14,
  Ge15}.
The study reported in this paper numerically demonstrates a nonlinear
dynamical mechanism for the formation of a lasing mode that adjusts
its pattern to a pumped area.

\begin{figure}[t!]
\centerline{\includegraphics[width=0.9\columnwidth]{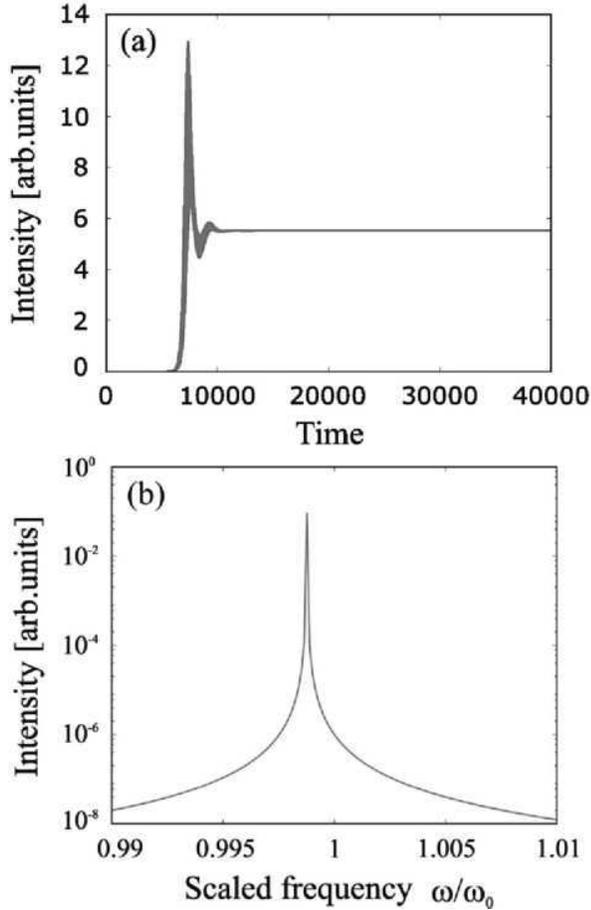}}
\caption{Results of the MB model simulation for the selective pumping
  case with $W_{\infty}$ $=$ $1.0$ $\times$ $10^{-3}$. (a) Time
  evolution of the total light intensity inside the cavity. (b) Power
  spectrum of the electric field for the stationary lasing regime. The
  peak frequency is around $\omega/\omega_0$ $=$ $0.9988$.}
\label{fig:MB_data}
\end{figure}

Figures \ref{fig:MB_states}(b) and \ref{fig:MB_data} show the results
of the selective pumping simulation when the pumping strength is set
at $W_{\infty}$ $=$ $1.0$ $\times$ $10^{-3}$, and the initial
distribution of the electric field $E_z$ is given by a Gaussian
distribution, as shown in Fig. \ref{fig:MB_states}(a).
For the selective pumping with $W_{\infty}$ $=$ $1.0$ $\times$
$10^{-3}$, the number of the modes that satisfy the positive linear
gain condition, Eq. (\ref{eq:positive_gain_condition}), turned out to
be 10, including the four nearly degenerate modes.
In Fig. \ref{fig:resonances_dist}, these 10 modes are indicated by
filled circles ($\bullet$).

Figure \ref{fig:MB_data}(a) shows the time evolution of the total
light intensity inside the cavity, where we can observe the formation
of a stationary lasing state after a transient.
Figure \ref{fig:MB_data}(b) shows the power spectrum of the time
series of the electric field at a certain point in the cavity taken in
the stationary regime.
From this power spectral data, we can confirm the single-mode lasing.
The corresponding time-averaged electric field intensity pattern is
shown in Fig. \ref{fig:MB_states}(b), where the average was taken over
the time interval of $T$ $\approx$ $4$ $\times 10^3$ $\times$
($2\pi/\omega_0$).
This pattern clearly shows that the stationary lasing state is a CW
rotating wave, which violates mirror symmetries with respect to both
the $x$ and $y$ axes.
Because the MB system with the triangle-orbit pumping pattern is
invariant under the transformation $x$ $\to$ $-x$, a CCW rotating wave
[obtained by reversing the $x$ axis in Fig. \ref{fig:MB_states}(b)] is
also a solution of the MB model.

\subsection{Interpretation of the Stationary Rotating-Wave States of the MB Model by the Resonant Modes}
\label{sect:interpretation}
It has been numerically demonstrated for the Schr\"odinger-Bloch (SB)
model that the frequencies of nearly degenerate modes can be locked as
a result of a nonlinear modal interaction, and the locking phenomenon
results in the appearance of asymmetric emission patterns
\cite{Harayama03, Sunada04, Sunada05, Shinohara05}.
The SB model is an approximation of the MB model, where the slowly
varying envelope approximation of the field variables is adopted for
reducing a numerical computation cost.

The power spectrum in Fig. \ref{fig:MB_data}(b) shows a single-mode
lasing, under the condition of preferential excitation of the four
nearly degenerate modes.
This result suggests that the locking occurs for these nearly
degenerate modes.
For confirming this, we computed the superposition of the
resonant-mode eigenfunctions.
First, we found that intensity localization along the upward-pointing
triangle orbit can be reproduced by taking the superpositions of two
different parity modes as follows:
\begin{align}
&\xi:=\psi_{ee}+\psi_{eo},\\
&\eta:=\psi_{oe}+\psi_{oo}.
\end{align}
\begin{figure}[t!]
\centerline{\includegraphics[width=1.0\columnwidth]{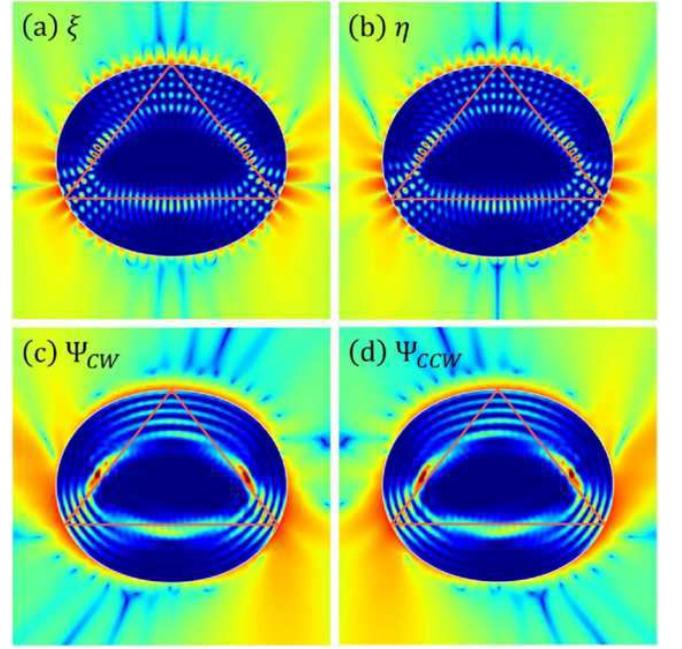}}
\caption{Intensity distributions of the superpositions of the
  resonant-mode wave functions. The triangle orbit is indicated by red
  lines, and the intensities outside the cavities are plotted in log
  scale. (a) $\xi$ $=$ $\psi_{ee}$ + $\psi_{eo}$. (b) $\eta$ $=$
  $\psi_{oe}$ + $\psi_{oo}$. (c) $\Psi_{CW}$ $=$ $\xi$ $+$ $i\eta$ $=$
  $(\psi_{ee}$ $+$ $\psi_{eo})$ $+$ $i(\psi_{oe}$ $+$
  $\psi_{oo})$. (d) $\Psi_{CCW}$ $=$ $\xi$ $-$ $i\eta$ $=$
  $(\psi_{ee}$ $+$ $\psi_{eo})$ $-$ $i(\psi_{oe}$ $+$ $\psi_{oo})$.}
\label{fig:wf_superpositions}
\end{figure}

We plot the intensity distributions of $\xi(x,y)$ and $\eta(x,y)$ in
Figs. \ref{fig:wf_superpositions}(a) and
\ref{fig:wf_superpositions}(b), respectively.
In the same manner, the intensity localization along the
downward-pointing triangle orbit can be obtained by $\xi'$ $:=$
$\psi_{ee}$ $-$ $\psi_{eo}$ and $\eta'$ $:=$ $\psi_{oe}$ $-$
$\psi_{oo}$.
Secondly, we found that the CW and CCW rotating-wave states can be
reproduced by taking the superpositions of $\xi(x,y)$ and $\eta(x,y)$
as follows:
\begin{align}
&\Psi_{CW}:=\xi+i\eta=(\psi_{ee}+\psi_{eo})+i(\psi_{oe}+\psi_{oo}),\\
&\Psi_{CCW}:=\xi-i\eta=(\psi_{ee}+\psi_{eo})-i(\psi_{oe}+\psi_{oo}).
\end{align}

The intensity distributions of $\Psi_{CW}(x,y)$ and $\Psi_{CCW}(x,y)$
are plotted in Figs. \ref{fig:wf_superpositions}(c) and
\ref{fig:wf_superpositions}(d), respectively.
Comparing Fig. \ref{fig:MB_states}(b) and
Fig. \ref{fig:wf_superpositions}(c), we can confirm that the CW
rotating-wave state of the MB model can be very well reproduced by the
superpositon of the four nearly degenerate double-triangle modes.

By performing simulations for various initial conditions and pumping
strengths, we numerically found that the CW and CCW rotating-wave
states are stable solutions of the MB model.
Because our selective pumping pattern is symmetric with respect to the
$y$ axis, when we prepare the initial field distributions symmetric
with respect to the $y$ axis, we obtained standing-wave stationary
states that obey the same symmetry.
We checked that these standing-wave stationary states can be
reproduced by the resonant-mode superpositions $\xi$ and $\eta$ shown
in Figs. \ref{fig:wf_superpositions}(a) and
\ref{fig:wf_superpositions}(b).
However, we numerically found that these standing-wave states are
dynamically unstable stationary solutions \cite{Harayama03}.

For the selective pumping, we found the lasing threshold to be
$W_{\infty}$ $\approx$ $2.8$ $\times$ $10^{-4}$.
Just above the threshold, we already observed the rotating-wave state
localized along the triangle orbit.
This appears to be reasonable, because the lasing mode needs to adjust
its pattern along the pumped area so that it can be excited.

The fact that the lasing mode at the threshold already localizes along
the triangle orbit yields the discrepancy between the actual lasing
threshold and the positive linear gain condition
Eq. (\ref{eq:positive_gain_condition}).
In the estimation of $\langle W_{\infty} \rangle$ by
Eq. (\ref{eq:W_inf}), the resonant-mode wave function $\psi(x,y)$ does
not localize only along the upward-pointing triangle orbit.
For the the resonant modes $\psi_{ee}$, $\psi_{eo}$, $\psi_{oe}$, and
$\psi_{oo}$, Eq. (\ref{eq:positive_gain_condition}) predicts the
positive linear gain thresholds to be $W_{\infty}$ $=$ 3.14 $\times$
$10^{-4}$, 3.32 $\times$ $10^{-4}$, 3.40 $\times$ $10^{-4}$, and 3.12
$\times$ $10^{-4}$, respectively, which are all larger than the actual
lasing threshold.
When we use $\xi(x,y)\pm i \eta(x,y)$ instead of $\psi(x,y)$ for the
estimation of $\langle W_{\infty} \rangle$ by Eq. (\ref{eq:W_inf}), we
found that the positive linear gain threshold is around 2.4 $\times$
$10^{-4}$, which is closer to the actual lasing threshold.
For a more accurate prediction of the lasing threshold, incorporating
the effect of the polarization term (thus the effect of the selective
pumping) in the linear Helmholtz equation \cite{Deych05, Ge10, Ge14,
  Ge15} is expected to be effective, which is, however, beyond the
scope of this paper.

\subsection{Uniform Pumping Simulation}
\begin{figure}[t!]
\centerline{\includegraphics[width=0.9\columnwidth]{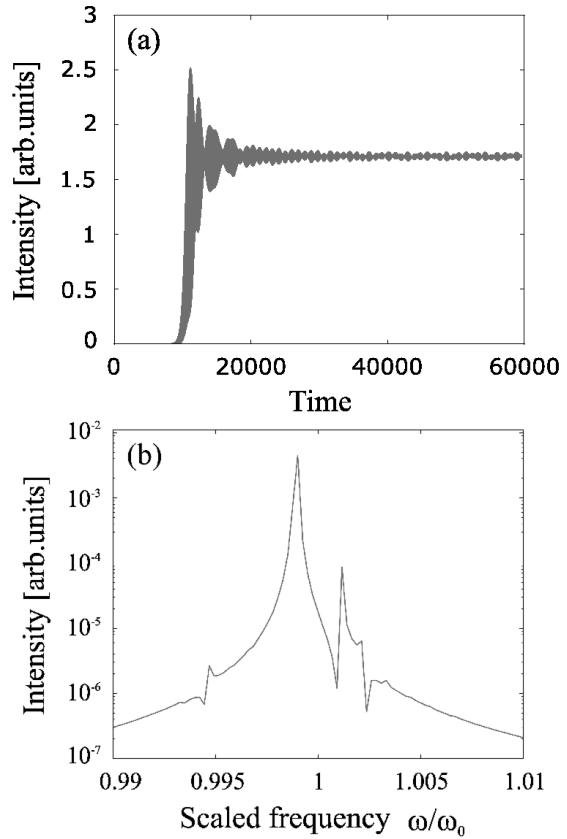}}
\caption{Results of the MB model simulation for the uniform pumping
  case with $W_{\infty}$ $=$ $3.0$ $\times$ $10^{-4}$. (a) Time
  evolution of the total light intensity inside the cavity. (b) Power
  spectrum of the electric field for the stationary lasing regime. The
  frequencies of the primary and secondary peaks are $\omega/\omega_0$
  $\approx$ $0.9990$ and $\omega/\omega_0$ $\approx$ $1.0012$,
  respectively.}
\label{fig:MB_unf_data}
\end{figure}
To compare with the selective pumping case, we carried out simulations
when the cavity is uniformly pumped.
We used the same parameter values as in the selective pumping case,
except for the pumping strength, which was set at $W_{\infty}$ $=$
$3.0$ $\times$ $10^{-4}$.
For this $W_{\infty}$ value, the number of the modes satisfying the
positive linear gain condition,
Eq. (\ref{eq:positive_gain_condition}), turned out to be 11, including
the four nearly degenerate double-triangle modes.

\begin{figure}[t!]
\centerline{\includegraphics[width=0.6\columnwidth]{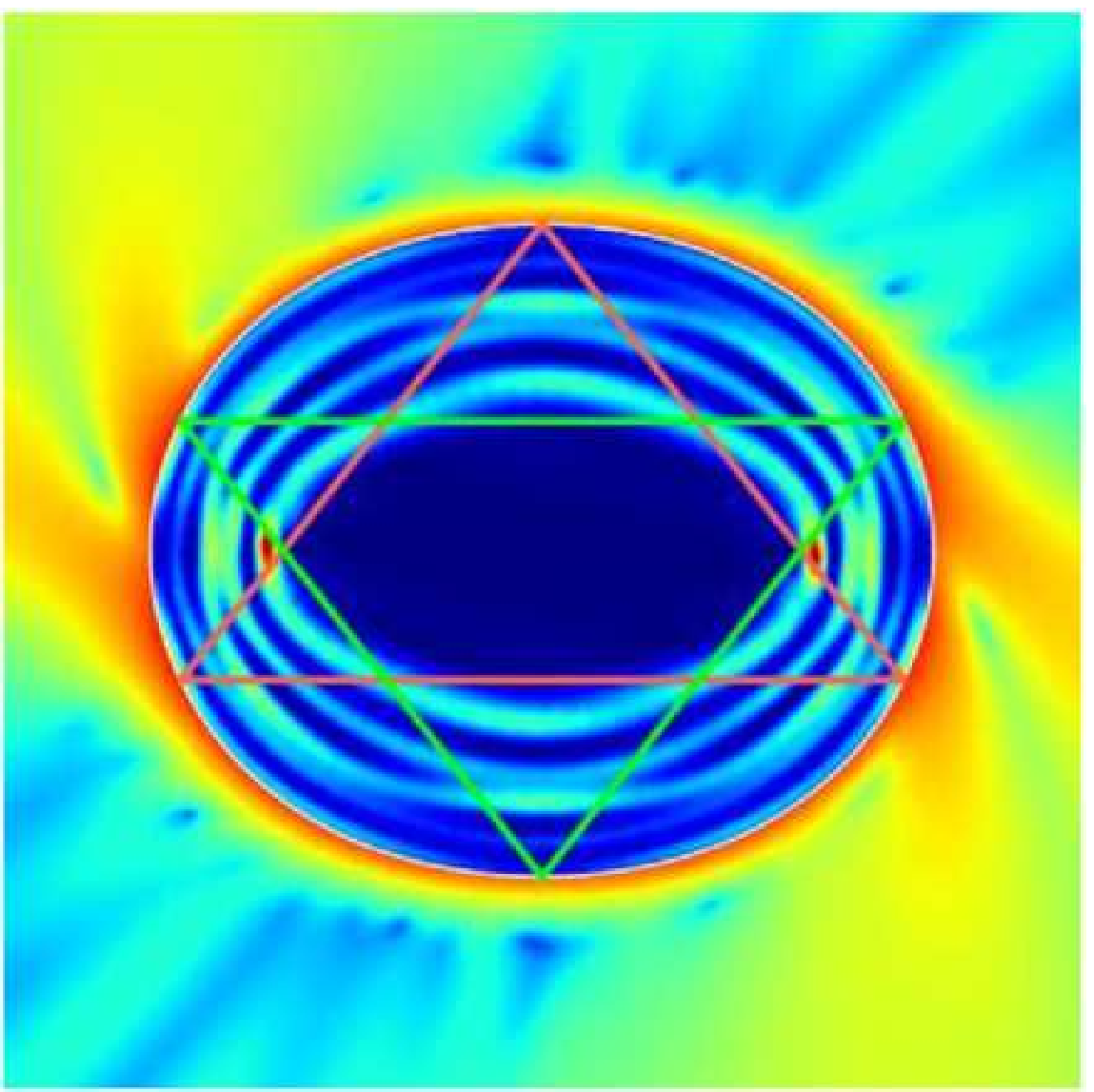}}
\caption{Time-averaged pattern of the stationary lasing state of the
  MB model for the uniform pumping case with $W_{\infty}$ $=$ $3.0$
  $\times$ $10^{-4}$. The intensity outside the cavity is plotted in
  log scale.}
\label{fig:MB_unf_wf}
\centerline{\includegraphics[width=1.0\columnwidth]{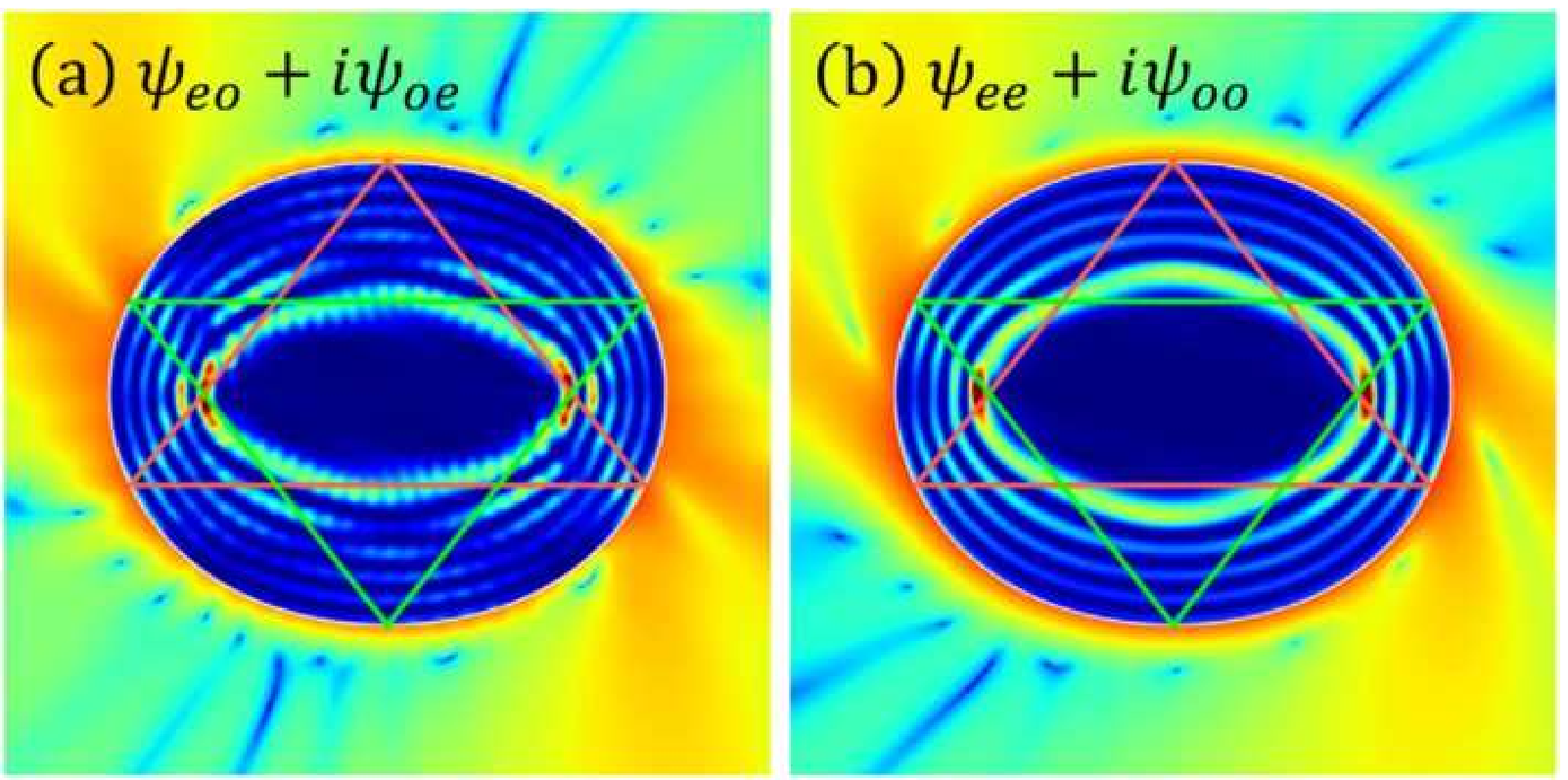}}
\caption{Intensity distribution of the superpositions of the
  resonant-mode wave functions. (a) $\psi_{eo}$ $+$
  $i\,\psi_{oe}$. (b) $\psi_{ee}$ $+$ $i\,\psi_{oo}$. The intensities
  outside the cavity are plotted in log scale.}
\label{fig:wf_superposition2}
\end{figure}

Figures \ref{fig:MB_unf_data} and \ref{fig:MB_unf_wf} show the results
of the MB model simulation with the uniform pumping.
Figures \ref{fig:MB_unf_data}(a) and \ref{fig:MB_unf_data}(b),
respectively, show the time evolution of the total light intensity
inside the cavity and the power spectrum of the electric field for the
stationary lasing regime.
In the spectral data, we can identify two dominant peaks at
$\omega/\omega_0$ $=$ $0.9990$ and $\omega/\omega_0$ $=$ $1.0012$ and
also observe their beat oscillation in the stationary regime of the
time-series data (we note that the ratio of the peak heights depends
on the position where the time-series data are acquired).
These data suggest that the stationary state consists mainly of the
two modes for $W_{\infty}$ $=$ $3.0$ $\times$ $10^{-4}$.
For higher pumping strengths (i.e., $W_{\infty}$ $\lesssim$ $1.0$
$\times$ $10^{-1}$), we did not observe the tendency for the two modes
to be locked.

We show in Fig. \ref{fig:MB_unf_wf} the time-averaged electric field
intensity pattern for the stationary lasing regime, where we can see a
CW rotating-wave state different from the one for the selective
pumping case.
By investigating the superposition of the resonant-mode wave
functions, we found that the pattern in Fig. \ref{fig:MB_unf_wf} can
be reproduced by $\psi_{eo}$ $+$ $i\psi_{oe}$ and $\psi_{ee}$ $+$
$i\psi_{oo}$.
Their patterns are shown in Fig. \ref{fig:wf_superposition2}(a) and
\ref{fig:wf_superposition2}(b), both of which show CW rotating-wave
patterns.
We note that the $eo$ and $oe$ modes have very close
eigenfrequencies, and so do the $ee$ and $oo$ modes (see the caption
of Fig. \ref{fig:wf_resonances} for their eigenfrequencies).
The scaled eigenfrequency $\mbox{Re}\,\omega/\omega_0$ of the former
pair is around 0.99908, whereas that for the latter is around 1.000828.
These values closely correspond to the peak positions in the spectral
data, which provides a strong support for our interpretation that each
of the two lasing modes is the locked state of a nearly degenerate
pair.

For the uniform pumping, we found the lasing threshold to be
$W_{\infty}$ $\approx$ $2.0$ $\times$ $10^{-4}$, which is smaller than
the threshold for the selective pumping.
This result seems to be natural, because the lasing modes are
different between the two cases, and for the selective pumping, the
double-triangle modes are partially pumped.

\section{CONCLUSION}
We numerically demonstrated the lasing of a triangle orbit mode in the
quadrupole-deformed microcavity laser with spatial selective pumping
along the periodic orbit.
We used the MB model to describe the nonlinear interaction between the
light field and a gain medium.
The nonlinear interaction is essential for the existence of the
triangle orbit lasing mode, as there is no corresponding resonant mode
for the passive cavity because of the $x$- and $y$-axes mirror
symmetries of the quadrupole cavity.
By a passive-cavity mode analysis, we concluded that the asymmetric
lasing mode can be interpreted as the locked state of the four nearly
degenerate modes associated with the double-triangle orbits.
In view of our theoretical results, the experimental study by Aung et
al. \cite{Aung15} is considered to be a very nice illustration of the
nonlinear dynamical effect on the formation of asymmetric lasing
modes.
Although our results presented here are for $\gamma_{\perp}$ $\approx$
$\gamma_{\parallel}$, we numerically confirmed that the locking
phenomena of the four nearly degenerate modes were similarly observed
for $\gamma_{\perp}$ $\gg$ $\gamma_{\parallel}$ (e.g.,
$\gamma_{\perp}$ $=$ $10^{-2}$ and $\gamma_{\parallel}$ $=$
$10^{-5}$).

We numerically confirmed that the CW and CCW rotating-wave triangle
orbit modes are stable solutions for the MB model, whereas
standing-wave ones are unstable.
It would be an interesting future problem to examine if a CW or CCW
rotating-wave stationary state can be obtained even when multiple
longitudinal modes are involved in lasing.

Another open issue is the detailed theoretical mechanism for the
nonlinear interaction of the four different parity resonant modes.
Our comparison between the selective and uniform pumping cases
revealed that the modal interaction mechanism does depend on the
pumping pattern.
Elucidating its mechanism would be useful for better controlling
lasing modes through selective pumping.

Regarding device optimization for threshold current reduction by the
selective pumping, it might be effective to adopt an active-passive
structure, where the non-pumped cavity area is made of a passive
material so as to suppress material absorption.

\section*{Funding}
Waseda University Grant for Special Research Projects (2017B-197).

\section*{Acknowledgment}
The authors thank Mr. Shunya Sekiguchi for his support in
ray-dynamical simulations. T. H. is grateful to Prof. Li Ge for
fruitful discussions.

\end{document}